\documentclass[twocolumn,letterpaper]{aastex61}
\usepackage{amsmath,amstext}
\usepackage[T1]{fontenc}
\usepackage{apjfonts} 
\usepackage[figure,figure*]{hypcap}

\shorttitle{VVV-J144321-611754}
\shortauthors{Baravalle, L. et al.}
\begin{document}

\title{The first galaxy cluster discovered by the VISTA Variables in the V\'ia L\'actea Survey}

 %\correspondingauthor{M. Victoria Alonso}
 %\email{m.v.alonso@gmail.com}

%\author{Author A}
%\affiliation{Affiliation 1}
%\affiliation{Affiliation 2}

\author{L. D. Baravalle}
\affiliation{Instituto de Astronom\'ia Te\'orica y Experimental, (IATE-CONICET), Laprida 854, C\'ordoba, Argentina}
\affiliation{Observatorio Astron\'omico de C\'ordoba, Universidad Nacional de C\'ordoba, Laprida 854, C\'ordoba, Argentina}
 
\author{J. L. Nilo Castell\'on}
\affiliation{Departamento de F\'isica y Astronom\'ia, Facultad de Ciencias, Universidad de La Serena, Av. Juan Cisternas 1200 Norte, La Serena, Chile}
\affiliation{Instituto de Investigaci\'on Multidisciplinario en Ciencia y Tecnolog\'ia, Universidad de La Serena. Avenida Juan Cisternas 1400, La Serena, Chile}

\author{M. V. Alonso}
\affiliation{Instituto de Astronom\'ia Te\'orica y Experimental, (IATE-CONICET), Laprida 854, C\'ordoba, Argentina}
\affiliation{Observatorio Astron\'omico de C\'ordoba, Universidad Nacional de C\'ordoba, Laprida 854, C\'ordoba, Argentina}

\author{J. D\'iaz Tello}
\affiliation{Preuniversitario UC, Pontificia Universidad Cat\'olica de Chile, Chile}
\affiliation{Instituto de Astronom\'ia, Sede Ensenada, Universidad Nacional Aut\'onoma de M\'exico, M\'exico}

\author{G. Damke}
\affiliation{AURA Observatory in Chile, Avenida Juan Cisternas 1500, La Serena, Chile} 
\affiliation{Instituto de Investigaci\'on Multidisciplinario en Ciencia y Tecnolog\'ia, Universidad de La Serena. Avenida Juan Cisternas 1400, La Serena, Chile}

\author{C. Valotto}
\affiliation{Instituto de Astronom\'ia Te\'orica y Experimental, (IATE-CONICET), Laprida 854, C\'ordoba, Argentina}
\affiliation{Observatorio Astron\'omico de C\'ordoba, Universidad Nacional de C\'ordoba, Laprida 854, C\'ordoba, Argentina}

\author{H. Cuevas Larenas}
\affiliation{Departamento de F\'isica y Astronom\'ia,  Facultad de Ciencias, Universidad de La Serena, 
Av. Juan Cisternas 1200 Norte, La Serena, Chile}
 
\author{B. S\'anchez}
\affiliation{Instituto de Astronom\'ia Te\'orica y Experimental, (IATE-CONICET), Laprida 854, C\'ordoba, Argentina}
\affiliation{Observatorio Astron\'omico de C\'ordoba, Universidad Nacional de C\'ordoba, Laprida 854, C\'ordoba, Argentina}
 
\author{M. de los R\'ios}
\affiliation{Instituto de Astronom\'ia Te\'orica y Experimental, (IATE-CONICET), Laprida 854, C\'ordoba, Argentina}
\affiliation{Observatorio Astron\'omico de C\'ordoba, Universidad Nacional de C\'ordoba, Laprida 854, C\'ordoba, Argentina}

\author{D. Minniti}
\affiliation{Millennium Institute of Astrophysics, Chile.}
\affiliation{Departamento de Ciencias F\'{\i}sicas, Universidad Andr\'es Bello, Rep\'ublica 220, Santiago, Chile.}
\affiliation{Vatican Observatory, Vatican City State V-00120, Italy}
    
\author{M. Dom\'inguez}
\affiliation{Instituto de Astronom\'ia Te\'orica y Experimental, (IATE-CONICET), Laprida 854, C\'ordoba, Argentina}
\affiliation{Observatorio Astron\'omico de C\'ordoba, Universidad Nacional de C\'ordoba, Laprida 854, C\'ordoba, Argentina}

\author{S. Gurovich}
\affiliation{Instituto de Astronom\'ia Te\'orica y Experimental, (IATE-CONICET), Laprida 854, C\'ordoba, Argentina}
\affiliation{Observatorio Astron\'omico de C\'ordoba, Universidad Nacional de C\'ordoba, Laprida 854, C\'ordoba, Argentina}
    
\author{R. Barb\'a}
\affiliation{Departamento de F\'isica y Astronom\'ia,  Facultad de Ciencias, Universidad de La Serena, 
Av. Juan Cisternas 1200 Norte, La Serena, Chile}

\author{M. Soto}
\affiliation{Instituto de Astronom\'ia y Ciencias Planetarias de Atacama, Universidad de Atacama, Copayapu 485, Copiap\'o, Chile}
\affiliation{Space Telescope Science Institute, 3700 San Martin Drive, Baltimore, MD 21218, USA}

\author{F. Milla Castro}
\affiliation{Departamento de F\'isica y Astronom\'ia,  Facultad de Ciencias, Universidad de La Serena, 
Av. Juan Cisternas 1200 Norte, La Serena, Chile}

\begin{abstract}

  We report the  first confirmed detection of the galaxy cluster VVV-J144321-611754 at very low latitudes (l = 315.836$^{\circ}$, b = -1.650$^{\circ}$) located in the tile d015 of the VISTA Variables in the V\'ia L\'actea (VVV) survey.  
  We defined the region of 30$\times$ 30 $arcmin^2$ centered in the brightest galaxy finding 25 galaxies. For these objects, extinction-corrected median colors of (H - K$_{s}$) = 0.34 $\pm$ 0.05 mag, (J - H) = 0.57 $\pm$ 0.08 mag and (J - K$_{s}$) = 0.87  $\pm$ 0.06 mag, and R$_{1/2}$ = 1.59 $\pm$ 0.16 $arcsec$; C = 3.01 $\pm$ 0.08; and Sersic index, n = 4.63 $\pm$ 0.39 were estimated.   They were visually confirmed showing characteristics of early-type galaxies in the near-IR images.
  An automatic clustering analysis performed in the whole 
tile found that the concentration of galaxies VVV-J144321-611754 is a real, compact concentration 
of early-type galaxies. Assuming a typical galaxy cluster with low X-ray luminosity, the photometric redshift of the
brightest galaxy is $z = $ 0.196 $\pm$ 0.025.  Follow-up near-IR spectroscopy with FLAMINGOS-2 at the Gemini-South telescope revealed that the two brighter cluster galaxies have typical spectra of early-type
galaxies and the estimated redshift for  the brightest galaxy VVV-J144321.06-611753.9 is $z =$ 0.234$\pm$0.022 and for VVV-J144319.02-611746.1 is $z =$ 0.232$\pm$0.019.  
Finally, these galaxies clearly follow the cluster Red Sequence in the rest-frame near-IR color--magnitude diagram with the slope similar to galaxy cluster at redshift of 0.2. These results are consistent
with the presence of a bona fide galaxy cluster beyond the Milky Way disk. 

\end{abstract}


%-------------Section 1 ------------------------------

\section{Introduction}

Extragalactic sources and large-scale structure behind the Milky Way (MW) are obscured by  dust and stellar crowding. 
The light of  these sources is dimmed more than $\sim$ 25\%  in the optical and  $\sim$ 10\% in the near-infrared wavelengths (Henning et al. 1998).   This effect is far larger close to the MW mid-plane because the dust is concentrated there. 
 
Several efforts have been made to study the distribution of 
galaxies behind the MW.  Kraan-Korteweg \& Lahav (2000) have reviewed the search of galaxies in the Zone of Avoidance (ZoA, |b| $< 20^{\circ}$) using optical, near-IR, far IR, radio and X-ray wavelengths. In the near-IR,  the 2MASS Redshift Survey (Huchra et al. 2012) 97.6\% complete to the limiting magnitude of K$_{s}$ = 11.75 mag was 
used to generate the catalog of groups of galaxies (Tully 2015) in the ZoA.  
At radio wavelengths, we can mention 
the Arecibo L-band Feed Array ZoA survey (McIntyre et al. 2015) and the blind 
21cm HI-line imaging survey on the Perseus-Pisces Supercluster filament 
crossing the ZoA (Ramatsoku et al. 2016). 
The results are complementary, near-IR selection favors the detection of 
early-type galaxies and blind HI surveys favor the  detection of late-type 
galaxies typically.

However, rich galaxy clusters  are traced by X-ray emmisions of the diffuse gas present in the systems.  A search for clusters of galaxies in the ZoA was carried out by
by Ebeling,  Mullis \& Tully (2002) using the ROSAT All Sky Survey
Bright Source Catalog (Ebeling et al. 1996; Voges et al. 1999).  There are only 13 galaxy clusters with measured redshifts in the literature at lower galactic latitudes (|b|$ < 2^{\circ}$): 1 cluster detected from XMM-Newton observations (Nevalainen et al. 2001); 2 clusters from the Clusters in the Zone of Avoidance project (Ebeling,  Mullis \& Tully 2002); 1 from the meta-catalogue of X-ray detected clusters of galaxies (Piffaretti et al. 2011); and 9  from the Planck catalogue of Sunyaev-Zeldovich sources
(Ma, Hinshaw \& Scott 2013).

The VISTA Variables in the V\'ia L\'actea (Minniti et al. 2010, hereafter VVV) is a public ESO photometric variability survey aimed mainly to study the stellar populations of the MW bulge (-10 $^{\circ}$$<$ $\ell$$<$ +10$^{\circ}$ and -10$^{\circ}$$<b <$+5$^{\circ}$) and MW disk  (-65$^{\circ}$$<$ $\ell$ $<$ -10$^{\circ}$ and -2$^{\circ}$ $< b <$ +2$^{\circ}$) in Z, Y, J, H and K$_s$ near-IR passbands. This survey has yielded the detection of new globular clusters (Minniti et al. 2011, 2017); new stellar open 
clusters (Borissova et al. 2011; Borissova et al. 2014; Barb\'a et al. 2015); the discovery of brown dwarfs (Beamin et al. 2013)  and other classes of variable stars. Moreover, data from the VVV survey have also been used for extragalactic studies. Am{\^o}res et al. (2012) reported 204 
candidate galaxies behind  the Galactic disk, in the direction of the Carina's spiral arm ($l \sim$ 285$^{\circ}$ to 290$^{\circ}$) based on visual inspection.  Coldwell et al. (2014) used near-IR photometry to confirm the existence of the previously X-ray detected galaxy cluster Suzaku J1759-3450 (Mori et al. 2013). Coldwell et al. (2014) found 15 galaxies within a  projected distance of 350 kpc from the central X-ray peak emission at a redshift of 0.13. More recently,  Baravalle et al. (2018) presented a method to search and characterize extragalactic sources detected from VVV image data. They derived photometric and morphological parameters of the sources by combining {\tt SExtractor+PSFEx}.

In this study, we use our photometric procedure to search for extragalactic sources and present the analysis for a galaxy cluster candidate found in the VVV tile d015.
VVV-J144321.06-611753.9 is the first confirmed galaxy cluster selected originally through VVV photometry.  
This paper is organized as follows: In section $\S$2, we describe the near-IR data and selection criteria, both for the VVV photometry and for the FLAMINGOS-2 spectroscopic data obtained at Gemini South. In section $\S$3, we present the cluster detection and analysis from four different methods, all used to confirm that VVV-J144321.06-611753.9 is a bona fide galaxy cluster. The  methods are (1) Automatic clustering analysis; (2) Spectral energy distribution modeling; (3) Spectroscopic redshift measurements; and (4) the Cluster Red Sequence.  In section $\S$4, we present a summary of the data analysis and main results. 
Throughout the paper, we use the following cosmological parameters:
H$_{0}$ = 70.4 kms$^{-1}$ Mpc$^{-1}$,
$\Omega_{M}$ = 0.272, and $\Omega_{\lambda}$ = 0.728 (Komatsu et al. 2011). \\

%----------------Section 2------------------------------------
\section{The data}

A concentration of extended objects was detected serendipitously in the search for new stellar cluster candidates in the VVV survey  (Barb\'a et al. 2015). 
VVV-J144321.06-611753.9 was found in the d015 MW disk tile centered at  J2000 $RA = 14h43m42.14s$, $Dec = -61^{\circ}40^{\prime}33.96^{\prime\prime}$ (l = 315.836$^{\circ}$, b = -1.650$^{\circ}$), which corresponds to the brightest object coordinates.

\subsection{VVV Near-IR photometry}

In order to detect and characterize the extragalactic sources in the tile d015, we applied the 
photometric methodology developed in Baravalle et al. (2018), which uses the combination
of ${\tt SExtractor\ v2.19.1}$ and ${\tt PSFEx\ v3.17}$.
The sources were selected using the  stellar index \textit{CLASS\_STAR}, the SPREAD\_MODEL ($\Phi$) 
parameter, the radius that contains 50\% of the total flux of an object 
(R$_{1/2}$), the concentration index (C) and the near-IR colors. 
The adopted criteria consider:  ${\tt CLASS\_STAR} < 0.3$; 1.0 $<$ R$_{1/2} < 5.0~arcsec$;
2.1 $<$ C $<$ 5; and $\Phi > 0.002$; and the colors  0.5 $<$ (J - K$_{s}$) $<$ 2.0 mag;
0.0 $<$ (J - H) $<$ 1.0 mag; and 0.0 $<$ (H - K$_{s}$) $<$ 2.0 mag with the color restriction 
(J - H) + 0.9 (H - K$_s$) $>$ 0.44 mag to minimize false detections.  Also, the sources with Z and Y detections should satisfy -0.3 $<$ (Y - J) $<$ 1.0 mag and -0.3 $<$ (Z - Y) $<$ 1.0 mag. 

\begin{figure}
\centering
\includegraphics[width=1.0\textwidth]{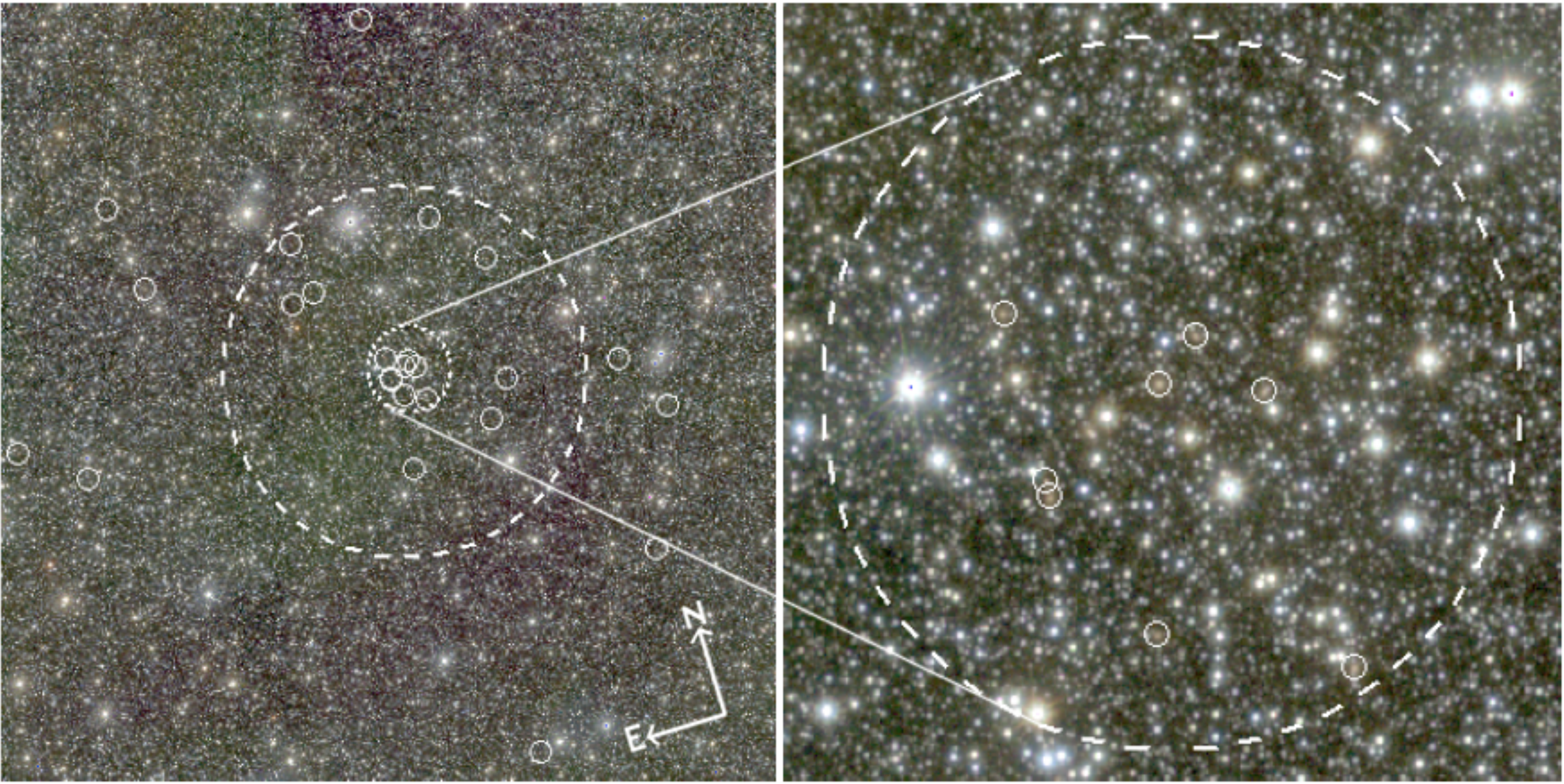}
\caption{Color  images of the concentration of galaxies showing the region of 30 $\times$ 30 arcmin$^2$ (left panel) and  the central parts of 3 $\times$ 3 arcmin$^2$ (right panel).  The 25 visually confirmed galaxies are surrounded with small circles. 
The circles with diameters of 3 and 7 arcmin are shown with dashed lines. 
}\label{field}
\end{figure}

Following these criteria, we detected 933 extragalactic sources  in the d015 VVV MW disk tile with 
24 located in the region of 30$\times$ 30 $arcmin^2$ centered at VVV-J144321.05-611753.9. 
Three objects were discarded because they are low surface brightness sources with strong
strong stellar contamination that could not be effectively corrected.  
The objects  VVV-J144243.0-611529.8, VVV-J144331.6-612133.1 and VVV-J144426.4-611101.3 have contamination by nearby stars and they were corrected using the procedure described in Baravalle et al. (2018).   Additionally, four sources that satisfied the above criteria but had slightly smaller R$_{1/2}$ than 1 arcsec, R$_{1/2}$, between 0.78 and 0.95 $arcsec$),
 were also included after a visual inspection.  Our final sample consists of 25 extragalactic sources, all of them visually confirmed to be galaxies.  They were all detected in the J, H and K$_{s}$ passbands and only four sources were also detected in the Z and Y VVV passbands. 
Figure~\ref{field} shows a cutout made from a composite (J, H and K$_{s}$) color image of the 
field showing the concentration of galaxies, centered at the brightest galaxy (VVV-J144321.05-611753.9).  The 25 galaxies of the concentration are marked by small circles.  The left panel corresponds to a  30$\times$30 $arcmin^2$ box identifying with dashed lines, two circles with diameters of 3 and 7 $arcmin$. The right panel shows the 3$\times$3 $arcmin^2$ central parts of the cluster candidate. 

Table~\ref{tab1} shows the photometric properties of the 25 galaxies. Column (1) gives the VVV 
identification; columns (2) and (3),  RA and Dec in J2000 coordinates; columns (4) to (8) show  
the  PSF Z, Y, J, H and K$_{s}$ magnitudes; columns (9) to (13) show the aperture magnitudes 
within a 2 $arcsec$ diameter, respectively; finally, columns (14) to (17) list the morphological 
parameters based on K$_{s}$-band photometry: the half-light radius; concentration index; the 
ellipticity; and the Sersic index.  The PSF magnitudes are estimates of the total K$_{s}$ 
magnitudes and  they range between 13.70 and 16.00 mag. All magnitudes were corrected for galactic 
extinction using the maps of Schlafly \& Finkbeiner (2011) and the relative extinctions of Catelan et al. (2011).  
The completeness was estimated for the tiles d010 and d115 in Baravalle et al. (2018, Figure 3) showing that 80 percent completeness is reached for sources fainter than 17 to 17.5 K$_{s}$ magnitudes.  We are reaching these magnitude limits in the tile d015, with a median  K$_{s}$ extinction value A$_{Ks}$ = 0.73 $\pm$ 0.01 mag.  In the studied region of 30$\times$30 $arcmin^2$, the extinction was not constant with median A$_{Ks}$ = 1.07 $\pm$ 0.04 mag:  
1.13 $\pm$ 0.05  mag in the central parts (radius smaller than 1.5 $arcmin$); 1.12 $\pm$ 0.07 mag for radius between 1.5 to 3.5 $arcmin$; and 0.93 $\pm$ 0.09 mag in the most external parts.     

The galaxies in the concentration are clearly small and red when compared against those in the whole tile.  
Extinction corrected median values of the photometric properties are:  PSF magnitudes, J = 15.50 $\pm$ 0.13 mag; H = 15.09 $\pm$ 0.14 mag; and K$_{s}$ = 14.64 $\pm$ 0.14 mag and colors, (H - K$_{s}$) = 0.34 $\pm$ 0.05 mag; (J - H) = 0.57 $\pm$ 0.08 mag; and (J - K$_{s}$) = 0.87  $\pm$ 0.06 mag.  The median morphological parameters are: R$_{1/2}$ = 1.59 $\pm$ 0.16 $arcsec$; C = 3.01 $\pm$ 0.08; $\epsilon$ = 0.30 $\pm$ 0.03 and Sersic index, n = 4.63 $\pm$ 0.39.  

\subsection{FLAMINGOS-2 IR spectroscopy}

We obtained  follow-up spectroscopic data using the near-infrared imaging spectrograph, FLAMINGOS-2 
at Gemini South, in  the Fast Turnaround (Program ID: GS-2016A-FT-18) observing mode. The 
observations were made during the nights of July 30 and August 02 of 2016. We collected long-slit spectra with 
the JH/HK grism and 0.72 arcsec slit.  This resulted in a peak resolution R$\sim$600, a dispersion of 6.55 \AA/pix and a total spectral coverage from 0.98 
to 1.80 $\mu$m. The projected on-sky slit length was 4.48 $arcmin$.  We observed the 8 brightest galaxies in the central part of the concentration. We set each pointing to include two
objects per slit, constraining the angular separation between each galaxy pair to be lower than 3 $arcmin$.  In addition, we optimized the slit position angle configuration or slit rotation to avoid bright stars for each setup. 
The objects have K$_{s}$ magnitudes between 13.7 to 14.8 mag and the exposure time was 8 $\times$ 120s to reach a signal-to-noise ratio $S/N > 3$ for each combined galaxy spectrum. Calibration frames were acquired together with telluric standard star spectra to remove telluric absorption features from the science data. 

The spectra were reduced using the IRAF/Gemini tasks and the Gemini FLAMINGOS-2 PyRAF data 
reduction package.  We performed the basic data reduction and combination of the 2-dimensional data with \textit{nsreduce} and \textit{nscombine} routines,
respectively.  
The arc-line identification and wavelength calibration steps were carried out with the 
\textit{nsfitcoords} and \textit{nstransform} routines, respectively. After obtaining the
two-dimensional spectra, the one-dimensional spectra were extracted via the \textit{nsextract} 
routine, to then be telluric-corrected following the Maiolino et al. (1996) procedure. Finally, 
the spectra were also flux calibrated using the photometric data available for the telluric 
standard star observed.  

%----------------Section 3------------------------------------
\section{Cluster Detection and Analysis}

In this section we present the photometric and spectroscopic results of the galaxy concentration 
VVV-J144321-611754.   We performed an automatic clustering analysis in the tile d015 and we 
obtained a photometric redshift estimate through the analysis of the spectral energy distribution. 
We determined the spectroscopic redshifts for the brightest objects and, finally, the 
cluster Red Sequence.

\subsection{Automatic clustering analysis}

We performed an automatic clustering analysis in order to gain insight about the nature of the galaxy clustering. This analysis aims to split the data into sets of associated objects with a joint conditional probability distribution. In our case, a mixture of simple Gaussian densities would separate the data into different groups by varying the number of free parameters, such as the number and shape of the Gaussians. The best model was obtained using the Bayesian Information Criteria, which penalizes those models with high number of free parameters.

\begin{figure}
\centering
\includegraphics[width=0.60\textwidth]{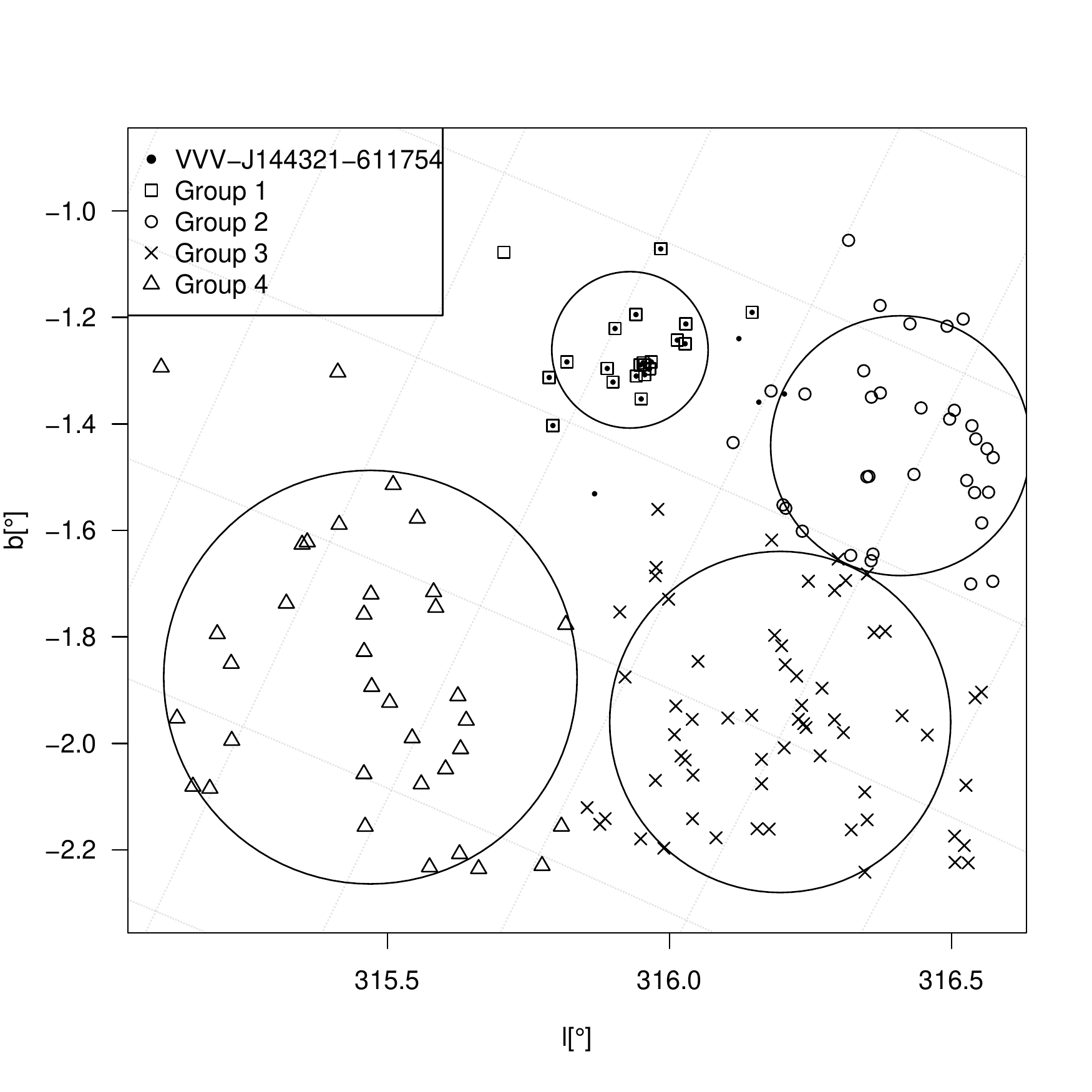}
\caption{Distribution of the extragalactic sources detected in the tile d015 in galactic coordinates.
The groups obtained in the clustering analysis are represented by different symbols and encircled by the estimated angular size. 
}\label{clustering}
\end{figure}

We used the public code Mclust (Scrucca et al. 2016) in the catalog of extragalactic sources found in the tile d015, applying the mixture of Gaussians to the sky coordinates, Right Ascension and Declination. In order to use galaxies with higher probability of belonging to a galaxy cluster, they should satisfy that: R$_{1/2} < $ 3 $arcsec$; n $>$ 2; and K$_{s}$ magnitudes fainter than 15.7 mag. We found that the best model has 4 spherically symmetric Gaussians associated to 4 groups with different circular angular areas. The group 1 identified with the concentration of galaxies VVV-J144321-611754 has the smallest angular radius of the sample, 0.138 degrees, whereas the other three groups have angular radii greater than 0.230 degrees.  Galaxies within this radius have a higher probability of belonging to the group. Figure~\ref{clustering} shows the distribution of the extragalactic sources of the tile d015 in galactic coordinates. The 25 galaxies of VVV-J144321-611754 are represented by small dots; the objects identified with the groups 1 to 4, with open squares, circles, crosses and triangles, respectively. The group angular sizes  are shown in Figure~\ref{clustering}.
We also took into account some photometric properties of the extragalactic sources, such as the extinction corrected colors, C and n to characterize the groups.  The median values for group 1 that contains  the concentration of galaxies are (J - K$_{s}$) = 0.84 $\pm$ 0.05 mag; (H - K$_{s}$) = 0.30  $\pm$ 0.04 mag; R$_{ 1/2}$ = 1.27 $\pm$ 0.09 $[arsec]$;  C = 3.01 $\pm$ 0.06; $\epsilon$ = 0.301 $\pm$ 0.03; n = 4.3 $\pm$ 0.3. The other three groups include bluer galaxies with C and n higher than 3.03 and 4.75, respectively.  Group 1 is formed with 24 extragalactic objects that were visually inspected; some of them are fainter
galaxies with nearby bright stars or located at the outer parts with lower group probability. This result suggests that the VVV-J144321-611754 is a real concentration of galaxies formed by early-type objects.

\subsection{Modeling the Spectral energy distribution}

EzGal (Mancone \& Gonzalez 2012) is a tool to model the evolution of the spectral energy distribution (SED)  of a stellar population that gives some derived properties such as the magnitude evolution as a function of redshift for different passbands.  This tool uses different stellar population synthesis (SPS) models including Bruzual \& Charlot (2003), the Maraston model (Maraston 2005), the BaSTI model (Percival et al. 2009), and the Flexible Stellar Population Synthesis (FSPS, Conroy, Gunn \& White 2009). The Bruzual \& Charlot (2003) model is the most commonly used, which computes the spectral evolution of stellar populations for different ages and  a wide range of  metallicities.  The Maraston model includes a detailed treatment of thermally pulsating asymptotic giant branch stars that dominate the infrared light of a young stellar population.  The BaSTI models  include a wide range of metallicities for both solar-scaled and $\alpha$-enhanced metallicities. The FSPS models are unique in dealing with important SPS inputs, such as, the initial mass function (IMF) or stellar evolution phases, treating them as free parameters, allowing the uncertainties introduced by various SPS inputs to be quantitatively measured.

The  K$_{s}$ model magnitudes were obtained using EzGal for Bruzual \& Charlot (2003) models assuming a formation redshift of z = 3; the Chabrier initial mass function; a solar metallicity; and a simple stellar population at different redshifts.  We used a standard normalization of K$_{s}$ absolute magnitude of -25.51 $\pm$ 0.09 (Brough et al. 2002) for redshifts higher than 0.1 assuming the cluster X-ray luminosity L$_{X}$(0.3 $--$ 3.5 keV) $< $ 1.9 $\times$ 10$^{44}$ erg s$^{-1}$.  In the evolution of K$_{s}$ magnitudes, the brightest galaxy in the concentration has an observed K$_{s}$ magnitude of 13.77 mag, which corresponds to a photometric redshift of 0.196 $\pm$ 0.025.  This value is also in agreement with the Stott et al. (2009) estimate based on the K$_{s}$ magnitude and color of the brightest cluster galaxy.  Therefore, all these results indicate that the photometric redshift of the galaxy concentration VVV-J144321-611754 is 0.196 $\pm$ 0.025.  

\subsection{Spectroscopic redshifts}

\begin{figure}
\centering
\includegraphics[width=77mm]{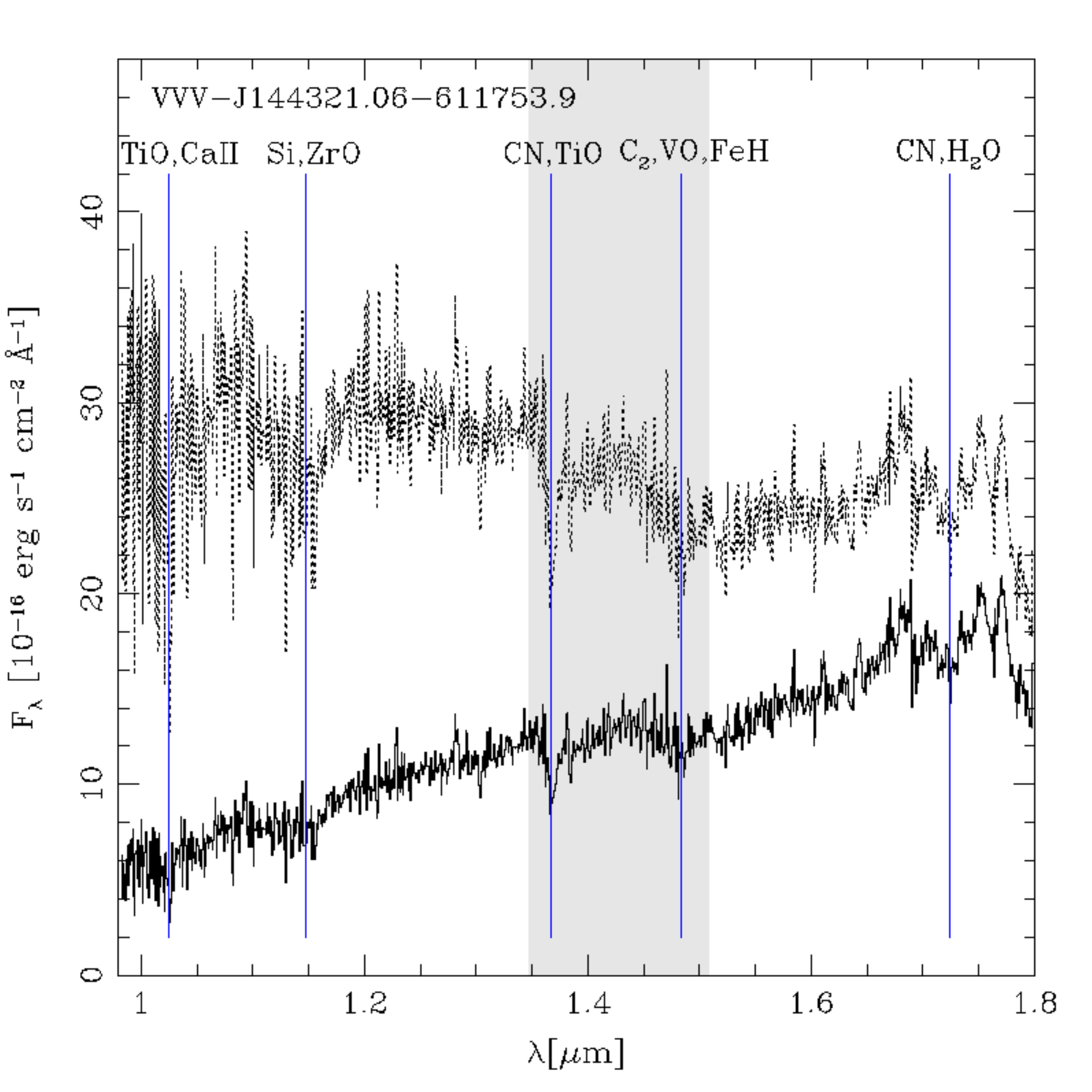}
\includegraphics[width=77mm]{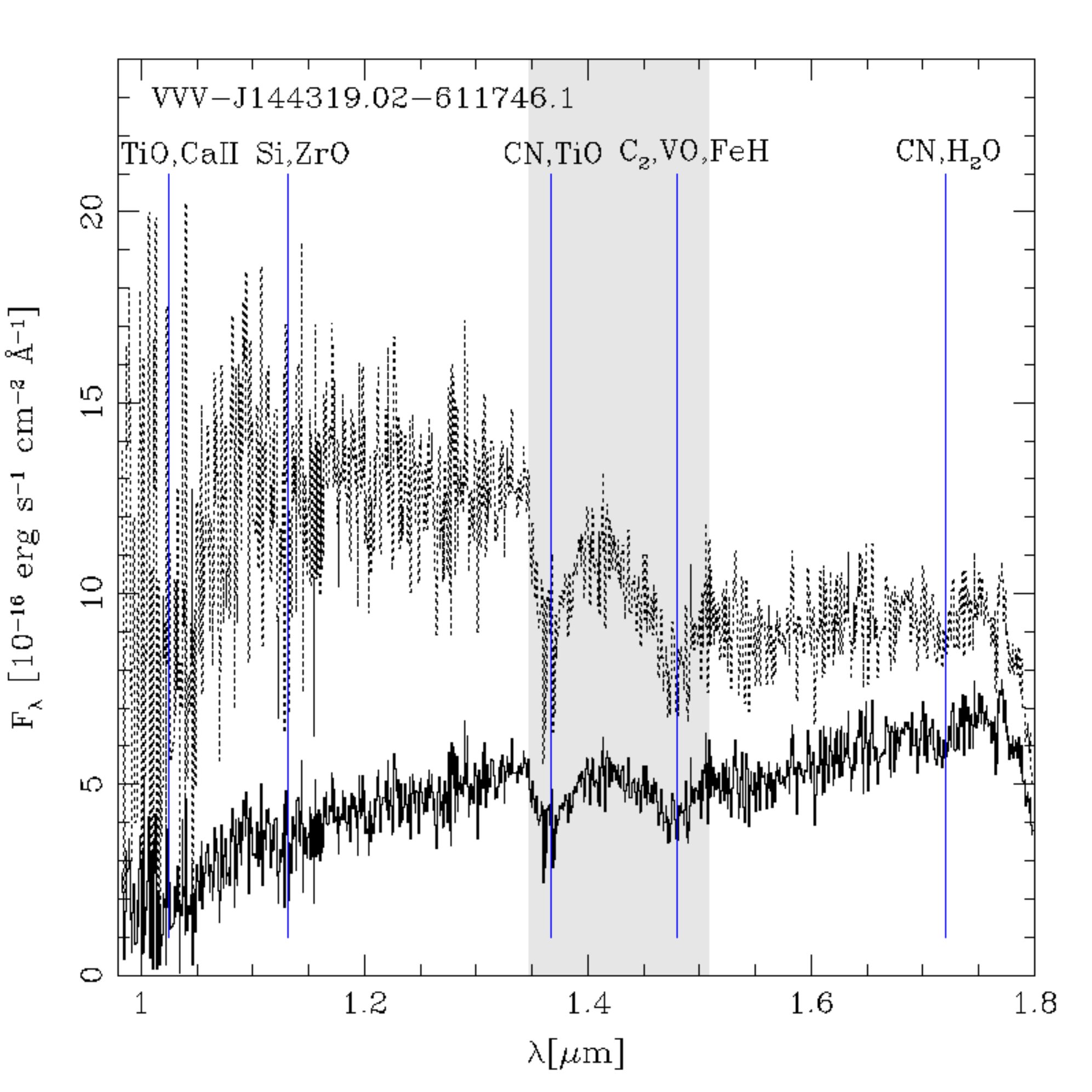}\\
\caption{Observer-frame galaxy spectra obtained in the long-slit mode of FLAMINGOS-2. The left panel shows the spectrum of VVV-J144321.06-611753.9 and the right panel, the spectrum of VVV-J144319.02-611746.1. The solid line represents the observed spectrum affected by the high dust extinction from the MW disk, while the dotted line shows the spectrum corrected by Galactic extinction. The vertical lines indicate the positions of the molecular bands and absorption lines identified in both spectra. The telluric absorption region has been shaded.
}\label{spec}
\end{figure}

The brighter galaxies VVV-J144321.06-611753.9 and VVV-J144319.02-611746.1 were observed in a single pointing.  The spectra were flux calibrated and then normalized to the photometry aperture fluxes, in order to be able to perform the dust reddening correction. The final spectra had $S/N ~$ 9.5 enough to obtain
a precise redshift estimation.  Figure~\ref{spec} shows the spectra of the two galaxies. The solid lines represent the observed spectra with the continuum  highly affected by the MW disk dust extinction, making the identification of the spectral features difficult.   The dotted lines represent the spectra corrected for absorption of Galactic dust.  The spectra were corrected by a simple model of dust reddening extinction in order to retrieve an hypothetical spectrum free of dust. We identified several molecular bands and absorption lines, such as 
the TiO $\lambda$0.85 $\mu$m and ZrO $\lambda$0.93 $\mu$m molecular bands overlapped with the CaII and Si absorption lines, respectively. Similarly, the CN and TiO bands at $\lambda$1.1 $\mu$m were identified, in addition to the blend of C$_{2}$, VO and FeH bands at $\lambda$1.2 $\mu$m. Finally, we recognized the slump of CN and H$_{2}$O $\lambda$1.4 $\mu$m bands. All these features, marked by vertical lines in the figure, are typical of old stellar populations (Rayner et al. 2009; Mason et al. 2015). The telluric absorption is shown as a shaded region.  The estimated redshift for VVV-J144321.06-611753.9 is $z=$ 0.234$\pm$0.022 and for VVV-J144319.02-611746.1 is $z=$ 0.232$\pm$0.019.  The relative large uncertainties in these estimates are mainly related to the use of molecular bands.  It was impossible to obtain redshifts for the fainter galaxies in the other pointings due to the lower $S/N$ of the spectra.

For the two brightest galaxies, the spectra include features typical of early-type galaxies.  
Also,  the good agreement between the two redshifts indicates that both galaxies likely belong to 
the same galaxy cluster. Also, for the brightest galaxy, the difference between the photometric and spectroscopic 
redshift estimates is about 0.044, and smaller than the uncertainties.
We use this spectroscopic redshift as a cluster redshift estimate.

\subsection{The Cluster Red Sequence}

In galaxy clusters, the early-type galaxies follow a tight relation in the Color-Magnitude 
Diagrams called the Red Sequence (Yee, Gladders \& L\'opez-Cruz 1999, hereafter RS).  
We applied the K-corrections to the K$_{s}$-band fluxes for the changes of the effective rest-
frame wavelengths.  We adopted the approximation to the computed observed colors of the 2MASS 
filters in the Vega system.  
Following the algorithm by Chilingarian et al. (2010), we used our estimated spectroscopic 
redshift of the galaxy concentration and the observed (J - K$_{s}$) color.

\begin{figure}
\centering
\includegraphics[width=1.0\textwidth]{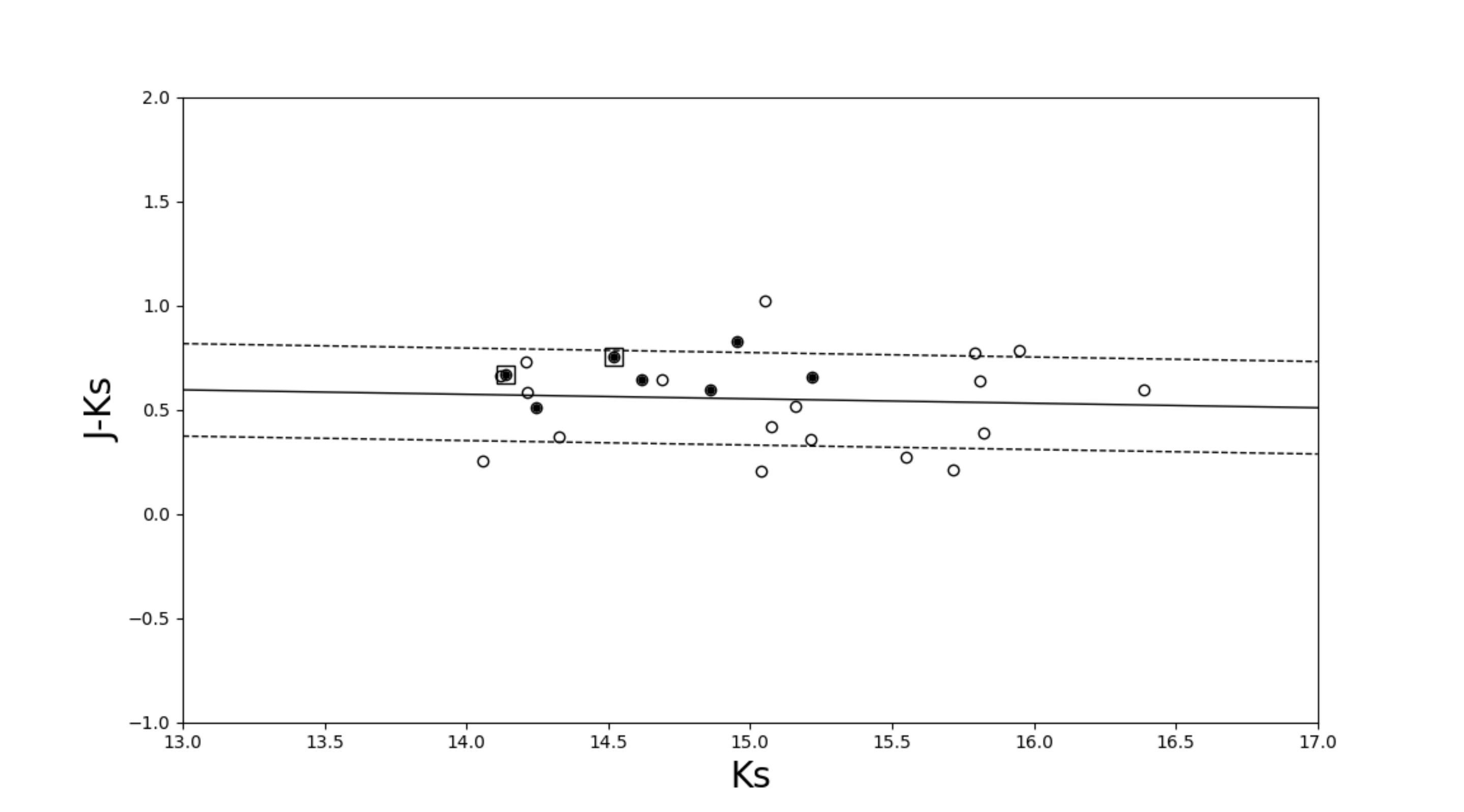}
\caption{ Near-IR color-magnitude diagram of galaxies in the concentration of galaxies VVV-J144321-611754.  The open circles represent the galaxies in the 30 $\times$ 30 $arcmin^2$ region; filled circles, the spectroscopic data; and open squares, the two objects with estimated redshifts.  The solid line represents the best RS linear fit and the dashed lines, the 1 $\sigma$ dispersion. 
}\label{rcs}
\end{figure}

Figure~\ref{rcs} shows the rest-frame near-IR (J - K$_{s}$) color -- K$_{s}$ magnitude diagram 
of the 25 galaxies in the 30$\times$30 $arcmin^2$ region.  
The objects with photometric data are represented by open circles; spectroscopic data by filled circles; and the two objects with estimated redshifts are represented by open squares.  The cluster RS shows substantial  scatter that might be related to the higher K$_{s}$ extinction values observed in the
 central parts of the cluster as compared to outer regions (Section $\S$2.1).  The best RS fit relation (J - K$_{s}$) = -0.02 $\pm$ 0.01 $\times$ K$_{s}$ +  0.87  $\pm$ 0.05  was obtained with the 22 galaxies that are also found to belong to the group 1 of the clustering analysis.  The solid line in the figure represents the best linear fit
 and the dashed lines, the 1 $\sigma$ dispersion.    Compared with Gladders et al. (1998) that found stable results using 3 $\sigma$ iteration, even this conservative RS has about 64\% of the galaxies in the central parts of the galaxy concentration.  We obtained similar results using bootstrapping analysis,  that supports the idea that these galaxies are part of a galaxy cluster. 
The cluster RS slope is in agreement with observed near-IR RS for the sample 
of Stott et al. (2009) and the slope evolution model by Bower et al. (2006) 
at z $\sim$ 0.2. 

Finally, we estimated the virial radius of the galaxy cluster at the derived redshift using the method of
Merch{\'a}n \& Zandivarez (2005).  We found R$_{vir}$ = 2.1 $\pm$  0.4 Mpc  for 
the galaxies in the group 1 of the clustering analysis, and R$_{vir}$ = 1.7 $\pm$ 0.2 Mpc for 
 the galaxies that follow  the RS within 1$\sigma$.   
The estimates agree with the typical virial radii of galaxy clusters
and they are comparable within the uncertainties. 

\section{Summary}

In this work, we report the first confirmed galaxy cluster, named  VVV-J144321-611754, originally detected 
in the tile d015 of the VVV survey.
We performed the photometric procedure in the tile as 
described in Baravalle et al. (2018) obtaining morphological and photometric parameters of 933 
extragalactic sources.  We defined a  
30$\times$ 30 $arcmin^2$ region centered in the brightest galaxy of the concentration, finding 25 
sources that were visually inspected and confirmed to be galaxies.  
For these galaxies, we obtain extinction corrected median near-IR colors of
(H - K$_{s}$) = 0.34 $\pm$ 0.05 mag, (J - H) = 0.57 $\pm$ 0.08 mag and 
(J - K$_{s}$) = 0.87  $\pm$ 0.06 mag.  
The median half-light radius is R$_{1/2}$ = 1.59 $\pm$ 0.16 $arcsec$; C = 3.01 $\pm$ 0.08; 
$\epsilon$ = 0.30 $\pm$ 0.03 and Sersic index, n = 4.63 $\pm$ 0.39.  
 All these morphological parameters are consistent with those of early-type  galaxies. 

An automatic clustering analysis found four groups in the tile, where the 
concentration of galaxies VVV-J144321-611754 is a real and the most compact concentration detected, composed of  
early-type 
galaxies. Assuming a typical galaxy cluster with low X-ray luminosity, we estimated the 
photometric redshift of 
the brightest galaxy to be $z =$ 0.196 $\pm$ 0.025.  
The IR spectra of the two brightest galaxies 
exhibit the typical features of early-type 
galaxies. The estimated redshift for VVV-J144321.06-611753.9 (the brightest galaxy) 
is $z=$ 
0.234$\pm$0.022 and for VVV-J144319.02-611746.1 is $z=$ 0.232$\pm$0.019. 
The estimated cluster 
redshift is $z=$ 0.234$\pm$0.022. The presence of the cluster RS is clear in the rest-frame 
color -- magnitude diagram of the studied 30 $\times$ 30 $arcmin^2$ region.   
Based on all these results, we conclude that the galaxy concentration VVV-J144321-611754 found in 
the VVV tile d015 is a bona fide galaxy cluster.

\acknowledgments
%\vskip 0.5cm
{\bf ACKNOWLEDGEMENTS}
%\vskip 0.5cm

This work was partially supported by grants from the Secretar\'ia de Ciencia y T\'ecnica (Secyt) of Universidad Nacional de C\'ordoba (UNC) and Consejo de Investigaciones Cient\'ificas y T\'ecnicas (CONICET). LDB acknowledges the financial  support for her PhD thesis from Secyt (UNC) and CONICET. JLNC is grateful for financial support received from the Programa de Incentivo a la Investigaci\'on Acad\'emica de la Direcci\'on de Investigaci\'on de la Universidad de La Serena (PIA-DIULS), Programa DIULS de Iniciaci\'on 
Cient\'ifica No. PI15142. JLNC also acknowledges the financial support from the GRANT PROGRAM No. FA9550-15-1-0167 of the Southern Office of Aerospace Research and Development (SOARD), a branch of the Air Force Office of the
Scientific Research's International Office of the United States (AFOSR/IO). FMC acknowledges the financial support from  Programa DIDULS  PT17145.

We gratefully acknowledge data from the ESO Public Survey program ID 179.B-2002 taken with the 
VISTA telescope, and products from the Cambridge Astronomical Survey Unit (CASU). DM is supported 
by the BASAL Center for Astrophysics and Associated Technologies (CATA) through grant AFB-170002, 
by the Ministry for the Economy, Development and Tourism, Programa Iniciativa Cient\'ifica Milenio 
grant IC120009, awarded to the Millennium Institute of Astrophysics (MAS), and by FONDECYT No.1170121.

%\facility{facility ID}
%\facilities{facility ID, facility ID, facility ID} 
%\software{Numpy}

%\bibliographystyle{yahapj}
%\bibliography{references}

%%%%%%%%%%%%%%%%%%%%%%%%%figures%%%%%%%%%%%%%%%%%%%%%%%%%

%%%%%%%%%%%%%%%%%%%%%%%%%tables%%%%%%%%%%%%%%%%%%%%%%%%%%
%% The values (usually only l,r and c) in the last part of
%% \begin{deluxetable}{} command tell LaTeX how many columns
%% there are and how to align them.
\begin{deluxetable}{ccccccccccccccccccc}

%% Rotate to a landscape orientation
\rotate

%% Over-ride the default font size
%% Use Default (12pt)

%% Use \tablewidth{?pt} to over-ride the default table width.
%% If you are unhappy with the default look at the end of the
%% *.log file to see what the default was set at before adjusting
%% this value.

%% This is the title of the table.
\caption{Photometric and morphological parameters of the galaxies in the galaxy concentration.}\label{tab1}

%% This command over-rides LaTeX's natural table count
%% and replaces it with this number.  LaTeX will increment 
%% all other tables after this table based on this number
\tablenum{1}

%% The \tablehead gives provides the column headers.  It
%% is currently set up so that the column labels are on the
%% top line and the units surrounded by ()s are in the 
%% bottom line.  You may add more header information by writing
%% another line between these lines. For each column that requries
%% extra information be sure to include a \colhead{text} command
%% and remember to end any extra lines with \\ and include the 
%% correct number of &s.
\tablehead{\colhead{ID} & \colhead{RA (J2000) } & \colhead{Dec (J2000)} & \colhead{Z} & \colhead{Y} & \colhead{J} & \colhead{H} & \colhead{K$_{s}$} & \colhead{Z$_{2\arcsec}$} & \colhead{Y$_{2\arcsec}$} & \colhead{J$_{2\arcsec}$} & \colhead{H$_{2\arcsec}$} & \colhead{K$_s$ $_{2\arcsec}$} & \colhead{R$_{1/2}$ } & \colhead{C} & \colhead{$\epsilon$} & \colhead{n} & \colhead{} & \colhead{} \\ 
%\colhead{} & \colhead{} & \colhead{} & \colhead{} & \colhead{} & \colhead{} & \colhead{} & \colhead{} & \colhead{} & \colhead{} & \colhead{} & \colhead{} & \colhead{} & \colhead{} & \colhead{} & \colhead{} & \colhead{} & \colhead{} & \colhead{}
} 
%% All data must appear between the \startdata and \enddata commands
\startdata
  VVV-J144217.23-612102.9 & 14:42:17.23 & -61:21:02.9 & 15.85   & 15.58   & 15.45 & 15.16 & 14.95 &  15.79  & 15.47   & 15.27 & 14.96 & 14.74 & 1.27 & 2.79 & 0.29 & 2.09 \\%9 1
  VVV-J144243.03-611529.8 & 14:42:43.03 & -61:15:29.8 & \nodata & \nodata & 14.87 & 14.16 & 13.79 & \nodata & \nodata & 14.57 & 13.83 & 13.53 & 2.16 & 3.80 & 0.16 & 3.45 \\%2 2
  VVV-J144252.33-611955.9 & 14:42:52.33 & -61:19:55.9 & \nodata & \nodata & 16.52 & 15.96 & 15.61 & \nodata & \nodata & 15.76 & 16.32 & 15.46 & 0.78 & 2.55 & 0.25 & 4.81 \\%13 3
   VVV-J144254.41-611308.4 & 14:42:54.41 & -61:13:08.4 & \nodata & \nodata & 16.63 & 15.96 & 15.53 & \nodata & \nodata & 16.5  & 15.84 & 15.39 & 1.10 & 2.8 & 0.34 & 4.63 \\%3 4
   VVV-J144302.30-612104.7 & 14:43:02.30 & -61:21:04.7 & \nodata & \nodata & 15.58 & 15.12 & 14.80 & \nodata & \nodata & 15.37 & 14.95 & 14.61 & 1.01 & 3.01 & 0.07 & 3.36 \\%1 5
   VVV-J144317.50-611807.6 & 14:43:17.50 & -61:18:07.6 & 14.64   & 14.71   & 14.62 & 14.30 & 14.06 & 14.65 & 14.62     & 14.40 & 14.07 & 13.85 & 1.79 & 3.39 & 0.34 & 3.39 \\%10 6
  VVV-J144318.67-611925.5 & 14:43:18.67 & -61:19:25.5 & \nodata & \nodata &  15.03 & 14.43 & 13.89 & \nodata & \nodata & 14.63 & 13.91 & 13.88 & 2.65 & 2.97 & 0.46 & 2.67 \\%17* 7
  VVV-J144319.03-611746.2 & 14:43:19.03 & -61:17:46.2 & \nodata & \nodata &  15.09 & 14.55 & 14.10 & \nodata & \nodata & 15.33 & 14.55 & 14.26 & 1.08 & 2.92 & 0.21 & 6.07 \\%14* 8
  VVV-J144321.05-611753.9 & 14:43:21.05 & -61:17:53.9 & \nodata &  \nodata & 14.60 & 14.07 & 13.73 & \nodata & \nodata & 14.56 & 13.95 & 13.60 & 2.33 & 3.49 & 0.15 & 7.15 \\%5* 9
 VVV-J144325.01-611854.7 & 14:43:25.01 & -61:18:54.7 & \nodata & \nodata  & 15.76 & 15.09 & 14.81 & \nodata & \nodata & 15.59 & 14.93 & 14.64 & 1.95 & 3.48 & 0.17 &  5.18 \\%4* 10
 VVV-J144325.32-611719.0 & 14:43:25.32 & -61:17:19.0 & \nodata & \nodata  & 15.16 & 14.57 & 14.22 & \nodata & \nodata & 15.02 & 14.48 & 14.09 & 1.73 & 3.25 & 0.30 & 4.63 \\%6* 11
 VVV-J144326.47-611804.6 & 14:43:26.47 & -61:18:04.6 & \nodata & \nodata  & 15.40 & 14.67 & 14.47 & \nodata & \nodata & 15.3  & 14.61 & 14.43 & 0.95 & 3.81 & 0.22 & 9.11 \\%16* 12
 VVV-J144326.58-611808.8 & 14:43:26.58 & -61:18:08.8 & \nodata & \nodata  & 15.67 & 14.95 & 14.53 & \nodata & \nodata & 15.35 & 14.65 & 14.19 & 1.59 & 2.89 & 0.45 & 2.39 \\%15* 13
  VVV-J144331.64-612133.1 & 14:43:31.64 & -61:21:33.1 & \nodata &  \nodata & 15.81 & 15.24 & 14.94 & \nodata & \nodata & 15.10 & 15.60 & 14.81 & 0.82 & 2.71 & 0.09 & 6.89 \\%11 14
  VVV-J144337.61-611355.6 & 14:43:37.61 & -61:13:55.6 & \nodata & \nodata  & 16.31 & 15.72 & 15.37 & \nodata & \nodata & 15.61 & 16.14 & 15.21 & 0.86 & 2.74 & 0.53 & 5.83 \\%12 15
  VVV-J144338.11-611153.2 & 14:43:38.11 & -61:11:53.2 & \nodata & \nodata  & 14.68 & 14.13 & 13.83 & \nodata & \nodata & 14.50 & 13.94 & 13.65 & 1.43 & 3.01 & 0.41 & 2.64 \\%8 16
  VVV-J144345.26-611356.6 & 14:43:45.26 & -61:13:56.6& \nodata  & \nodata  & 15.87 & 15.57 & 15.54 & \nodata & \nodata & 15.81 & 15.39 & 15.23 & 1.92 & 3.10 & 0.44 & 2.48 \\%7 17
  VVV-J144208.98-612325.1 & 14:42:08.98 & -61:23:25.1  & \nodata & \nodata & 16.55 & 15.86 & 15.37 & \nodata & \nodata & 16.31 & 15.67 & 15.22 & 1.04 & 2.87 & 0.48 & 6.62 \\  %18
  VVV-J144426.62-610744.4 & 14:44:26.62 & -61:07:44.4  & \nodata & \nodata & 15.11 & 14.51 & 14.29 & \nodata & \nodata & 15.12 & 14.4  & 14.19  & 1.42 & 3.46 & 0.08 & 5.90 \\ %19
VVV-J144230.74-612811.3 &     14:42:30.74 & -61:28:11.3  & 15.21 & 15.86     & 15.39 & 15.11 & 14.77 & 15.91 & 15.86     & 15.59 & 15.34 & 14.97  & 2.3  & 2.39 & 0.29 & 2.66 \\
VVV-J144331.39-613319.8 &     14:43:31.39 & -61:33:19.8  & \nodata & \nodata & 16.99 & 16.28 & 16.0  & \nodata & \nodata & 16.75 & 16.18 & 15.88  & 1.33 & 2.83 & 0.39 & 4.09 \\
VVV-J144248.72-610518.6 &     14:42:48.72 & -61:05:18.6  & \nodata & \nodata & 15.81 & 15.12 & 14.64 & \nodata & \nodata & 15.91 & 14.92 & 14.515 & 2.71 & 3.29 & 0.65 & 4.31  \\
VVV-J144426.38-611101.3 &     14:44:26.38 & -61:11:01.3  & \nodata & \nodata & 15.5  & 13.9  & 13.9  & \nodata & \nodata & 14.07 & 13.49 & 13.7   & 2.44 & 3.48 & 0.21 & 1.74  \\
VVV-J144507.49-611636.8 &     14:45:07.49 & -61:16:36.8  & \nodata & \nodata & 16.05 & 15.71 & 15.41 & \nodata & \nodata & 16.11 & 15.58 & 15.19  & 4.09 & 3.43 & 0.68 & 4.67  \\
VVV-J144524.77-611437.3 &     14:45:24.77 & -61:14:37.3  & 15.86 & 15.34     & 14.7  & 14.06 & 13.72 & 15.86 & 15.26     & 14.6 & 14.17 & 13.641 & 3.06 & 3.76 & 0.44 & 7.37  \\
\enddata
%% Include any \tablenotetext{key}{text}, \tablerefs{ref list},
%% or \tablecomments{text} between the \enddata and 
%% \end{deluxetable} commands
%% No \tablecomments indicated
%% No \tablerefs indicated
\end{deluxetable}

%\appendix

\end{document}